\documentstyle[12pt]{article}
\begin{document}
\begin{center}

{\large
{\bf Spin-charge decoupling and  orthofermi  quantum statistics}
}

\vskip 1.2cm

{\bf A. K. Mishra $^{*\dagger}$ }

\vskip 0.5cm

{\it {$^*$ Max-Planck Institute for Physics of Complex Systems,  
Nothnitzer Str. 38, D-01187 Dresden, Germany, and  \\
$^{\dagger}$ Institute of Mathematical Sciences,
CIT Campus, Madras - 600 113, India\\
e-mail: mishra@imsc.ernet.in}}

\end{center}

\vskip 1.2cm

\baselineskip=20pt
\centerline{\bf Abstract}

\vskip .5cm

\noindent Currently Gutzwiller projection technique and nested Bethe ansatz are 
two main methods used to handle electronic systems in the  $U$ infinity limit.
We demonstrate that these two approaches  describe two distinct physical
systems. In the nested  Bethe ansatz solutions, there is a decoupling between 
the spin and charge degrees of freedom.  Such a decoupling  is absent in
the Gutzwiller projection technique. Whereas in the Gutzwiller approach, the
usual antisymmetry of space and spin coordinates is maintained, we show
that the Bethe ansatz wave function is compatible with a new form of
quantum statistics, viz., orthofermi statistics. In this statistics,
the wave function is antisymmetric in spatial coordinates alone. This
feature ultimately leads to spin-charge decoupling.

\newpage

We had earlier envisaged a quantum system of spin-half particles
obeying a new exclusion principle, viz., an orbital state should
not contain more than one particle, whether spin up or spin down.$^{1}$
This modified exclusion principle is more restrictive than 
Pauli's principle, which allows two electrons having opposite
spin to occupy the same orbital state. When the Coulomb interaction $U$
between two such electrons tends to infinity, the resulting
system naturally satisfies the new exclusion principle (NEP).

\vskip .5cm

In the present paper, we consider the physical consequences
of such a singular potential on a system of electrons. In particular,
we show  that in the limit
of $U \rightarrow \infty$, there exists a possibility that the
antisymmetry under the simultaneous exchange of spatial and spin
coordinates in an electronic wave function may be violated.
Instead, the antisymmetry is valid only with respect to the
spatial coordinates, whereas no symmetry is $ {\it a~ priori}$ imposed on
the spin component of a multiparticle wave function. 
Next we elucidate how this new  symmetry of the wave function
ultimately leads to  (i) a new form of quantum statistics,
namely, orthofermi statistics, $^{1-3}$ and (ii) a novel concept of
spin-charge decoupling, proposed in the context of high
temperature superconductivity.$^{4-7}$ We also  
provide  a critical comparison of our results with the existing 
algebraic approaches to  the $U$ infinity problem.

\vskip .5cm

Both Pauli's exclusion principle and the NEP can be  formulated in terms of  particle
creation operator $c^{\dagger}$. Given an orbital index $i$, and
spin indices $\sigma$, $\bar{\sigma}$, the former is expressed
as

$$
c^{\dagger}_{i\alpha} c^{\dagger}_{i\alpha} \ = \ 0 ,~~~~ (\alpha \ =  \
\sigma~ or~ \bar{\sigma}).  \eqno(1) 
$$
The more exclusive NEP satisfies

$$
c^{\dagger}_{i\alpha} c^{\dagger}_{i\beta} \ = \ 0 ,~~~~ (\alpha, \beta \ = \ 
\sigma ~or~ \bar{\sigma}).  \eqno(2) 
$$
The relation (2) completely takes into account the local effect of the intraorbital
infinite Coulomb potential, and leads to an alteration  in the conventional
fermionic Fock or state space. 

\vskip .5cm

It is pertinent here to ask whether (i) the $U$ infinity constraint has any
possible nonlocal manifestation, and (ii) it is  possible to describe such a system
consistently through appropriate commutation relations involving
particle creation and annihilation operators? Concerning the first query, 
we note that the earlier-mentioned modified symmetry property of the  
wave function reflects
the nonlocal consequence of the $U$ infinity limit. 
The answer to the second question is also in the affirmative. In fact
consistent with the NEP, two independent sets of commutation
relations, valid for any spatial dimension, can be constructed.$^{1}$ These
relations are identical when only a single orbital state 
is considered, but differ when different orbital indices are involved.

\vskip .5cm

The first set of the commutation relations (CRI) is 
$$
c_{i \alpha} c_{j\beta}   +   (1-\delta_{ij}) c_{j\beta} c_{i\alpha} 
  =   0, \eqno(3)
$$

$$
c_{i\alpha} c^{\dagger}_{j\beta}  + (1 - \delta_{ij}) c^{\dagger}_{j\beta} c_{i\alpha}
\ = \ \delta_{ij}\delta_{\alpha \beta}
(1 - \sum_{\gamma} c^{\dagger}_{i\gamma} c_{i\gamma}). \eqno(4)
$$

Here Latin and Greek indices, respectively, specify the
spatial and  spin coordinates.

\vskip .5cm

The second set of commutation relations (CRII) is 

$$
c_{i\alpha} c_{j\beta} + c_{j\alpha} c_{i\beta} \ = \ 0,
 \eqno(5)
$$

$$
c_{i\alpha} c^{\dagger}_{j\beta} \ = \ \delta_{\alpha\beta}
(\delta_{ij} - \sum_{\gamma} c^{\dagger}_{j\gamma} c_{i\gamma}).\eqno(6)
$$

\vskip .5cm

Using either set of commutation relations independently, it can be
shown that the only permissible states associated with an 
orbital index $i$ are $| 0 \rangle,~  |1_{i\sigma} \rangle$,  and $
|1_{i\bar{\sigma}} \rangle .$ The states $\{|1_{i\alpha}
1_{i\beta} \rangle \}$ are always null states. Thus both sets of
commutation relations are compatible with the NEP. Now a
system of electrons subject to  the $U \rightarrow \infty$
constraint cannot be described through two distinct types of
commutation relations. But before taking up this important
aspect of the problem, we consider salient features as well
as the critical differences between the two algebras. This step
would  also help us in determining the correct set of commutation
relations.

\vskip .5cm

The CRI is not invariant under the unitary
transformation
$$
d_{i\alpha} \ = \ \sum_{j} U_{ij} c_{j\alpha} ; \quad
UU^{\dagger} = U^{\dagger}U = 1. \eqno(7)
$$
As a consequence, it is not preserved under a change in
representation. This contrasts with the usual fermionic
anticommutation relations
$$
\{f_{i\alpha} , f_{j\beta}\} \ = \ 0 ; \quad
\{f_{i\alpha} , f^{\dagger}_{j\beta}\} \ = \ \delta_{ij}
\delta_{\alpha \beta}, \eqno(8)
$$
which are representation invariant.

\vskip .5cm

The CRI is invariant under the phase transformation
$$
e_{i\alpha} \ = \ e^{i\phi_{i\alpha}} c_{i\alpha}. \eqno(9)
$$
Therefore, the particle number operator $N_{i\alpha}$ exists.
Next, the commutation relation (3) signifies the
antisymmetry when both spatial and spin coordinates are
simultaneously exchanged. We note that both these features
are valid for the usual fermions too.

\vskip .5cm

The overall antisymmetry prevents the spin-charge decoupling
when the number of particles $N$ exceeds 2. This is true for
the particles obeying CRI as well as for canonical fermions.
In fact for an $N$ particle system having $M$ down spins,
the spatial and spin part of the wave function, respectively,
satisfy conjugate symmetries $[2^K 1^{N-2K}]$ and $[N-K, K],~
0 \le K \le M$.$^{8}$  In this case, the associated wave function 
cannot be represented as a product of the Slater determinant corresponding to spinless  
fermions multiplied by  a spin wave function, or a superposition of such states,
unless $N \le 2$ or $K = 0$.

\vskip .5cm

The  CRII  is  invariant under the unitary transformation
(7), and hence it  is representation invariant. On the other hand,
the relations in CRII are not invariant under the phase transformation (9)
involving both indices $i$ and $\alpha$. These are invariant
only with respect to the following phase transformations :
$$
h_{i\alpha} = e^{i\phi_{i}} c_{i\alpha} ; \quad
l_{i\alpha} = e^{i\phi_\alpha} c_{i\alpha}. \eqno(10)
$$
Consequently only number operators $N_i$ and $N_\alpha$
exist, but not the number operator $N_{i\alpha}$. We note
here that it is the relation (5) in CRII that decouples the
spatial and spin coordinates. It is this decoupling that 
prevents us from mapping $i$ and $\alpha$ to a single index - 
a mapping which is possible for usual fermions and in CRI.
Therefore, we cannot
define $N_{i\alpha}$ with composite index $i\alpha$. Only
the number operators $N_i$ and $N_\alpha$ are allowed.

\vskip .5cm

The commutation relation (5) implies that a state vector is
antisymmetric only when the spatial indices $i$ and $j$ are
exchanged, whereas states having different permutations of
$\sigma$ and $\bar{\sigma}$ are independent, orthonormal
states. It may be noted here that as far as the spin variables
are concerned, a similar situation arises in Greenberg's
infinite statistics.$^{9,10}$ The only difference is that in 
infinite statistics, the range of index $\alpha$ is
unrestricted, whereas in the present context, $\alpha$ can
take only two values, viz., $\sigma$ and $\bar{\sigma}$.

\vskip .5cm

To summarize, the quanta characterized by CRII satisfy
Fermi-Dirac statistics with respect to spatial coordinates and
infinite statistics with regard to the spin variable. Even though
no ${\it a~ priori}$ symmetry restriction is imposed on the spin
variables, it is always possible to construct states that  are an
eigenfunction of a given spin Hamiltonian. To achieve this,
one has to take an appropriate superposition of states constructed through
strings of creation operators acting on the unique vacuum
state $|0 \rangle $.

\vskip .5cm

We have termed the statistics associated with the CRII as
orthofermi statistics. This represents an
instance wherein quanta having composite statistical
character (viz., indices belonging to different classes exhibit uncorrelated
permutation properties) is proposed.

\vskip .5cm

The above discussion makes it clear that the  CRI and CRII
describe two fundamentally distinct physical systems.

\vskip .5cm

Algebraic relations similar to the CRI have been reported 
earlier in the context of Gutzwiller projection $^{11}$ and
Hubbard operators.$^{12-14}$ But  problems exist with these
earlier constructions. The algebra satisfied by the Gutzwiller
projection operator is not closed, as shown in Ref. 1. On the other hand,
Hubbard operators are local in nature,$^{14}$ and additional
postulates are needed to obtain their multisite relations.$^1$
These problems do not arise in the CRI. 

\vskip .5cm

In the context of CRII, the quantum mechanical problem
associated with a system of electrons in one dimension,
mutually interacting with delta-function potential having
weight $U$, can be mentioned. This problem has been exactly
solved using the nested Bethe ansatz (NBA) by Yang.$^{15}$
The NBA uses Bethe ansatz twice - once for the charge
(or spatial) degree of freedom, and thereafter for the spin
component. The solution for the discrete version of this
problem, namely the one-dimensional Hubbard Hamiltonian
 
$$
H \ = \ t \sum_{<ij>\sigma} c_{i\sigma}^{\dagger} c_{j\sigma}  
  +  U \sum_{i} n_{i\sigma} n_{i \bar{\sigma}}  \eqno(11)
$$

\noindent has been given by Lieb and Wu.$^{16}$ The $U \rightarrow \infty$ limit of 
the 1D Hubbard Hamiltonian has been  considered by Ogata and 
Shiba.$^{4,5}$  In their work, the $U$ is taken to be  infinity
while calculating the spatial component of the wave function.
The ensuing wave function is a slater determinant describing 
noninteracting spinless fermions. As for the spin part, $U$ is not
exactly infinity in the sense that terms  up to the order of $t^2/U$  are retained.
This is equivalent to effectively replacing the Hubbard
Hamiltonian by the $t-J$ Hamiltonian where $J = 4t^2/U.
^{4-6}$  The complete wave function is a product of a Slater 
determinant involving only the spatial variables multiplied by a spin
wave function which is the exact solution of the 1D Heisenberg
Hamiltonian. It may be mentioned here that if $U$ is put exactly as infinity
in the sense that $J = 0$, all possible $2^N$ spin configurations become
degenerate. Retaining a  $J$ as $4t^2/U$ removes this degeneracy.$^6$ However, the symmetry
properties of the spin degree of freedom remains the same whether $J = 0$ or is
finite.

\vskip .5cm

It has been also mentioned in the literature  that instead of using
complicated NBA, this wave function can be obtained directly  
through simple physical arguments.$^{5,6}$
The three main characteristics  of the wave function 
are as follows.

\vskip .5cm

\begin{enumerate}
\item Decoupling occurs between space and spin components.
\item The spatial part of the wave function is antisymmetric when two
spatial coordinates are exchanged. It vanishes when any two
coordinates coincide.
\item Because of the factorization or  decoupling 
between the space and  spin
parts, it is no longer possible to specify whether a particle with a given
spatial coordinate  also has a  definite spin 
$\sigma$ or $\bar{\sigma}$, and vice versa. As a result, the number operator
$N_{i\alpha}$ with composite index $i\alpha$ cannot be defined.
\end{enumerate}

\vskip .5cm

These features of the wave function are valid both for the half-filled
and less than half-filled cases. In the later case, the wave function is
given as a superposition of such charge-spin decoupled states.
That the factorization of the wave function
constitutes a remarkable result has been earlier highlighted by Fulde.$^{6}$

\vskip .5cm

We have shown earlier $^{1}$ that the orthofermions
characterized by CRII  possess all of the above three
characteristics.

\vskip .5cm

We may also clarify why in the $U \rightarrow \infty$ limit, the symmetry
property of the wave function gets altered only in the NBA, and not in the 
Gutzwiller projection technique. In the conventional approach, the symmetry property 
of the wave function
(and hence the statistics of the system) 
is ${\it a ~ priori}$  postulated. Starting with a given eigenfunction, the corresponding wave function
having a particular symmetry is  obtained through appropriate usage of permutation operators.
For example, for a three-particle boson system, we can start with the 
state $\psi(1,2,3)$, and get the symmetric state using the following symmetrizer:
$$
\psi_s(1,2,3)  \ = \ \{I + P(12) + P(13)  + P(23) + P(123) + P(132) \}\psi(1,2,3). \eqno(12)
$$
We note that in this process of generating the wave function of a particular symmetry
type, permutation operators are never scaled by any dynamical variable.

\vskip .5cm

In the NBA, one starts  with the wave function in a particular ordered configuration.$^{4,15,16}$
Next, in order to generate the wave function (of a required symmetry type)
over the entire configuration space, we
use permutation operators and suitable boundary conditions.
In this process, the permutation operators get scaled 
by the dynamical variables $t$ and  $U$. And this is quite a nontrivial feature.
We note that this scaling is absent when
$U \ = \ 0$, and the corresponding situation is similar to the one depicted in Eq. (12).
On the other hand, when we take the $U$ infinity limit, the permutation operators
linking the one sector to another (or exchanging the particles residing in the
adjacent sectors), according to the desired symmetry requirement, 
become redundant.
To make these points more explicit, we consider the wave function
given in Refs. 4 and 16. The amplitude in the wave function, when down spins
are located at the sites $x_1, ....,x_M$ and up spins at $x_{M+1}, .....,x_N$,
is given as 

$$
f(x_1, .....,x_N) \ = \ \sum_P [Q,P] exp (i \sum_{j = 1}^N k_{Pj} x_{Qj}),
\eqno(13)
$$
\noindent where $ P = (P1, P2,....., PN)$ and $ Q = (Q1, Q2,....., QN)$ are two permutations
of $(1, 2,....., N)$ and $f$ is given in the sector $x_{Q1} < x_{O2} < .....
 < x_{QN}$. The { $[Q,P]$ }  are determined by the relation

$$
[Q,P] \ = \ Y_{nm}^{i,i+1} [Q,P^{'}],
\eqno(14)
$$
\noindent where $ P  = (P1, P2,..., Pi = m, P(i+1) = n,....., PN)$ and
 $ P^{'}  = (P1, P2,....., n, m,....., PN)$, 

$$
Y_{nm}^{i,i+1} \ = \ \frac {\displaystyle {P_{i,i+1} - x_{nm}}}
{\displaystyle { 1 + x_{nm}}}
\eqno(15)
$$

\noindent and

$$
x_{nm} \ = \ i(U/2)/(t~ sink_n - t~ sink_m).
\eqno(16)
$$
\noindent $P_{i,i+1}$ is a permutation operator for the interchange between
$Qi$ and $Q(i+1)$ and for an $N$  particle system, it admits an  
appropriate $N!\times N!$  matrix representation of the symmetry group $S_N.  
^{8,15}$ It may be noted here that the 
amplitude in the wave function is in fact  the
wave function in coordinate representation.$^{15}$

In Eq. (15), if we take  antisymmetric representation for $P_{i,i+1}$, then
$P_{i,i+1} [Q,P^{'}] = -[Q,P^{'}]$  and hence $[Q,P] = -[Q,P^{'}]$ follows from
Eq. (14). Note that this result is not valid for any other representation
for $P_{i,i+1}$ when $U$ is either zero or finite. 
But in case $U$ is taken to be infinity in Eq. (15),
$[Q,P] = -[Q,P^{'}]$  always holds true, irrespective of the 
representation of $P_{i,i+1}$. It is this peculiar  limiting behavior
of the operator $Y$ that is responsible for the spin-charge decoupling
in the $U$ infinity limit, and subsequently gives rise to spinless
fermions.

\vskip .5cm

In the Gutzwiller projection technique, on the other hand, 
we start with  antisymmetric wave functions. These wave functions are
constructed analogous to Eq. (12), but one uses an antisymmetrizer in  place
of a symmetrizer operator. In fact the wave function is a Slater
determinant built from the Bloch states of electrons on a lattice, and the  
$U$ infinity constraint is then implemented through 
the projection operator $\Pi_i (1 - n_{i\sigma} n_{i\bar{\sigma}})$.$^{6}$
Note that in this process of constructing the wave function, the
permutation operators never get scaled by $U$ or $t$. Also  the original
fermionic antisymmetry now remains intact.

\vskip .5cm

It would be instructive here to consider the particular case of
NBA when the $U$ infinity limit is taken for both spatial and spin components. It has been
often argued  that in this situation, electrons cannot
exchange their positions within the 1D chain.$^{5-6}$  But a closer
examination reveals that the antisymmetry associated with the
spatial part of the wave function allows us to exchange the
spatial coordinates of the electrons. Similarly, a finite $J$
term implies that spin coordinates can also be permuted. However
when $J=0~ ( U=\infty)$, it may appear that the spin sequence gets
frozen to its initial order. If this is true, then the
decoupled spin component of the system would satisfy the `null statistics'
.$^{17}$  Also the earlier statement that the electrons cannot
exchange their positions within the chain has to be modified to read,
`the  spatial coordinates can be exchanged in the 1D chain,
but not the spin coordinates.' 

\vskip .5cm

Irrespective of the above
discussion, we ought to demand that even when $J=0$, the
wave function is still an eigenfunction of the total $S^2$ and
$S_z$ operators. Consequently, for a given set of spatial coordinates
and with $S_z$ being $(N - 2M)/2$, all the spin configurations,
classified according to $\{[N - K, K],~ 0 \le K \le M\}$ symmetries,
are accessible. On the other hand, in the case of frozen spin order, only
one spin configuration is allowed for a given set of spatial
coordinates and a given value of $S_z$.

\vskip .5cm

We have already highlighted the important distinction between CRI
and CRII. Next, the $U=\infty$ case of the Hubbard Hamiltonian, i.e., 
$$
H \ = \ t \sum_{<ij>\sigma} c_{i\sigma}^{\dagger} c_{j\sigma} \eqno(17)
$$

\noindent can be taken as a typical example to 
show how these commutation relations  lead to different types of
dynamical evolution. We consider the time evolution equation for
$c_{i\sigma}$, which is needed to evaluate single particle Green's function or
correlation function. In the case where $c^{\dagger}, c$ satisfy CRII,
$$
i\dot c_{i\sigma} \ = \ t \sum_j c_{j\sigma}, \eqno(18)
$$
where $\dot c_{i\sigma}$ is the time derivative of $ c_{i\sigma}$.
The presence of  single annihilation operators  on the right hand side implies
that the system remains in the original operator space 
$\{c_{i\sigma}\}$ and its dual $\{c_{i\sigma}^{\dagger}\}$.
Consequently, the Hamiltonian is exactly solvable, a feature
compatible with the NBA case.

\vskip .5cm

If $c^{\dagger}, c$ obey the CRI, 
$$
i \dot c_{i\sigma} \ = \ t \sum_j \{c_{j\sigma}
- c^{\dagger}_{i\bar{\sigma}} c_{i\bar{\sigma}} c_{j\sigma} + c^{\dagger}_{i\bar{\sigma}}
c_{i\sigma} c_{j\bar{\sigma}} \}\,, \eqno(19)
$$
that is, the system evolves to an enlarged operator product space.
Alternatively stated, the presence of a triple operator product in
the above equation implies no closure in the equation of the motion
chain. Thus an exact solution cannot be obtained in this case;
appropriate truncations or projections are needed to arrive at
approximate solutions.$^{14}$

\vskip .5cm

By seeking the mutual consistency between  (i) the conjugate
permutation symmetries associated with the spatial and spin
components of the wave function of usual fermions, and (ii) the $U$
infinity constraint, it is possible to provide a more transparent
physical argument for the validity of orthofermi statistics and Bethe ansatz solutions 
and the ensuing  spin-charge decoupling. For
definiteness we consider the continuum case and take $N > 2$. The $U$ infinity
constraint demands that all multiparticle wave functions must
vanish when the spatial coordinates of any two particles coincide
$$
\psi(x_1, x_2 \ldots , x_i \ldots x_j \ldots x_N, \alpha_1,
\alpha_2 \ldots , \alpha_N) \rightarrow 0 \quad {\rm as} \quad
x_i \rightarrow x_j \quad {\rm for \ any} \quad \{i,j\},
\eqno(20)
$$
otherwise the  energy of the system diverges. This is true for any 
number of dimensions. As noted earlier, for a
$N$ electron system having M  down spins, the respective
permutation symmetries for spatial and spin parts of the
wave function are $^{8,15}$

$$
[2^K 1^{N-2K}] \quad {\rm and} \quad [N-K,K]~;~ 0 \le K \le M\,.
$$

\noindent Now to be consistent with the $U$ infinity case, we should retain
only those wave functions that  have a node when any two position
coordinates coincide, and remove all others that  do not
have such nodes. Accordingly, the only permissible spatial
wave function corresponds to $[1^N]$ symmetry, i.e., the  $K=0$ case. All
other wave functions corresponding to finite $K$ and $N > 2$ do not possess
required nodes. Also the spatial and spin coordinates are not decoupled
for the wave functions characterized by mixed-symmetry representations.
If we demand total antisymmetry including the
spatial and spin parts, only allowed spin configuration
corresponds to the conjugate symmetry $[N]$. Though one is able
to achieve  spin-charge decoupling in this case, the dimension
of the state  space stands greatly reduced. Now only symmetric
spin configuration is permissible. If we stipulate 
that all eigenfunction of  $S^2$ and $S_z$
are permissible, then every  $2^N$ spin configuration,
classified according to $\{[N-K,K]\}$ symmetries, is  allowed. But all
these spin configurations are coupled with the same  spatial
configuration $[1^N]$, and not with the conjugate spatial
configuration $[2^K 1^{N-2K}]$. It is this $[1^N] \times
[N-K,K]$ symmetry of the permissible wave function that  leads to
the spin-charge decoupling.
The CRII for the orthofermions obviously reflects this property of
the wave function at kinematic level. 

\vskip .5cm
To conclude, we have considered  two sets of commutation
relations (CRI and CRII) compatible with the `no 
double occupancy in a single orbital'  constraint for spin-half particles.
The subsequent analysis  brings out the distinctive features
of CRI and CRII, both at the level of kinematics and dynamics,
and highlights the possibility of a  violation of fermionic
antisymmetry in the present context.
Since Gutzwiller projection technique is closer to CRI, and the 
nested Bethe ansatz to CRII, we conclude that these two widely used 
approaches  to model the $U$
infinity constraint lead to quite different physical consequences.
In fact, they describe two distinct physical systems. We have finally provided 
the reasons why electrons in $U$ infinity limit may behave like orthofermions described by the CRII leading to a spin-charge decoupling, 
and have supported our
analysis through  a comparison  with the exact NBA
solutions.

\vskip .5cm
I am  grateful to Professor  P. Fulde for a critical
reading of the manuscript and for suggesting improvements. I am also indebted to
him  for providing valuable insights and many enlightening discussions during 
my visit to MPI-PKS. I thank Professors. G. Rajasekarn and  G. Baskaran for discussions
and various comments.

\end{document}